# Alignment of the photoelectron spectroscopy beamline at NSRL

LI Chao-Yang(李朝阳)    PAN Han-Bin(潘海斌)    WEN Shen(魏珅)    PAN Cong-Yuan(潘从元)

AN Ning(安宁)    DU Xue-Wei(杜学维)    ZHU Jun-Fa(朱俊发）WANG Qiu-Ping(王秋平)[*]

National Synchrotron Radiation Laboratory (NSRL), University of Science & Technology of China, Hefei, Anhui 230029, China

**Abstract**: The photoelectron spectroscopy beamline at National Synchrotron Radiation Laboratory (NSRL) is equipped with a spherical grating monochromator with the included angle of 174°. Three gratings with line density of 200, 700 and 1200 lines/mm are used to cover the energy region from 60 eV to 1000 eV. After several years' operation, the spectral resolution and flux throughput were deteriorated, realignment is necessary to improve the performance. First, the wavelength scanning mechanism, the optical components position and the exit slit guide direction are aligned according to the design value. Second, the gratings are checked by Atomic Force Microscopy (AFM). And then the gas absorption spectrum is measured to optimize the focusing condition of the monochromator. The spectral resolving power E/ΔE is recovered to the designed value of 1000@244eV. The flux at the end station for the 200 lines/mm grating is about $10^{10}$ photons/sec/200mA, which is in accordance with the design. The photon flux for the 700 lines/mm grating is about $5\times10^{8}$ photons/sec/200mA, which is lower than expected. This poor flux throughput may be caused by carbon contamination on the optical components. The 1200 lines/mm grating has roughness much higher than expected so the diffraction efficiency is too low to detect any signal. A new grating would be ordered. After the alignment, the beamline has significant performance improvements in both the resolving power and the flux throughput for 200 and 700 lines/mm gratings and is provided to users.

## 1. Introduction

The photoelectron spectroscopy beamline is designed for research on electronic states of solid surface using photoemission spectroscopy [1]. A typical spherical grating monochromator (SGM) with the included angle of 174° is used in this beamline [2-5]. Three gratings, G1, G2 and G3, with line density 200, 700 and 1200 lines/mm are used to cover the whole energy region from 60 eV to 1000eV. The designed resolving power of this beamline is 1000@244eV, and the designed flux throughput at the end station is $5\times10^{9}$ photons/sec/200mA. The optical layout and critical components of this beamline is shown in Fig.1.

---

[*] Corresponding author. Tel: +86 551 63602119

 *Email address: qiuping@ustc.edu.cn*

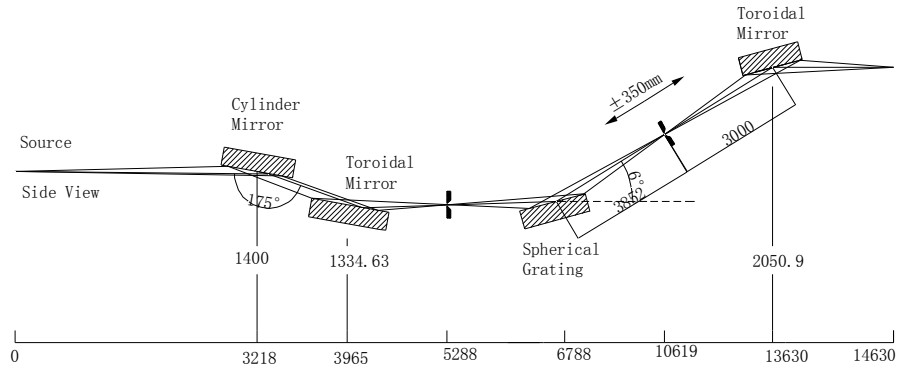

Fig. 1 Optical layout of the SGM beamline in NSRL

After several years' operation, performance of this beamline deteriorated and can't meet the experiments requirement. The Au 4f photoelectron spectrum before alignment in fig.2 showed the state of this beamline. The spectrum is measured with 450 eV photons. The measured line width (FWHM) of Au $4f_{5/2}$ and $4f_{7/2}$ are 1.21 eV and 1.3 eV. Considering the width of the energy analyzer is 0.1eV, and the natural linewidth is 0.54 eV, the resolution is estimated about 1.18eV, which corresponds to a resolving power of 380.

Therefore, realignment is needed to improve the performance of the beamline. This paper describes the previous status and performance of the beamline, realignment procedures, and measured results [6-7].

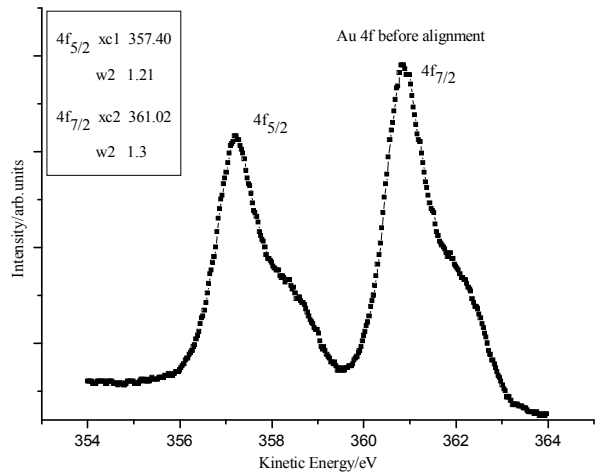

Fig. 2 Au 4f photoelectron spectrum before alignment

## 2. Realignment

Due to several years' operation, the optical component positions in the beamline have changed from their original values, and the surfaces of these optical components have been carbon contaminated. These reasons make the beamline flux and spectral resolution worse. All the critical optical components need to be aligned to recover its performance.

### 2.1 Critical components position check

All the critical position of components in the spherical grating monochromator (includes entrance and exit slits, gratings and the exit slit linear guide) are checked by the theodolite and the automatic level. It is found that the entrance slit, the exit slit and its linear guide have errors related to the grating rulings and grating center. If the slit opening is not parallel to the grating ruling direction, the resolution will be affected. About 1mrad of this error for the entrance and the exit slit. If the exit slit linear guide doesn't pass through the grating center, the included angle will be changed during slit scanning on it. This will cause the wavelength calibration error, focusing error, and the beam moving at the sample. All errors found are adjusted and set to the design value.

### 2.2 Resolution of the wavelength scanning mechanism

The resolution of the wavelength scanning mechanism has to be better than 0.5 arcsec to meet the energy resolving power, which is 1000@244eV. The length of the grating sine bar is 500 mm, so the resolution of the linear stage is calculated to be better than 0.001 mm [8]. But the resolution of the linear stage used is just 0.003mm. So, a new linear stage is used to replace the old one. The resolution of this new one is 0.0002mm.

### 2.3 Roughness of gratings

Roughness of the optical components will influence the flux throughput of beamline, so roughness of these grating is tested by ATF. Roughness of G3 is about 5nm, and the groove shape is damaged, so the diffraction efficiency of G3 is low. A new grating with line density of 1200 lines/mm will be ordered.

After initial alignment, the beamline is baked to ultra-high vacuum. It is ready to test the spectral resolution and calibrate the wavelength of the monochromator by the gas ionization chamber which is installed at the beamline just before the experimental station [9-10].

## 3. Performance after realignment

### 3.1 Resolving power

In order to estimate the spectral resolution of the monochromator and to calibrate the wavelengths, the photon absorption spectra for excitation of the inner-shell electron into the unoccupied states in the argon, krypton and nitrogen gases were measured by a gas ionization chamber [8-9]. Both entrance and exit slits' widths are set to 50 μm. First, the spectral resolution of grating with the 700 lines/mm is tested. Fig.3 (a) shows the photoionization spectrum of argon. The gas pressure in the ionization chamber is 7.7 Pa. FWHM at 244.39 eV is 0.16 eV. By taking a natural linewidth of $\Gamma$=0.114 eV, the resolving power E/ΔE at 244.39 eV is 2200. Fig.3 (b) shows the photoionization spectrum of nitrogen. The gas pressure is 3.8 Pa. FWHM at 406eV is about 0.4 eV. Assuming a natural linewidth of $\Gamma$=0.132 eV, the resolving power at 406 eV is 1000.

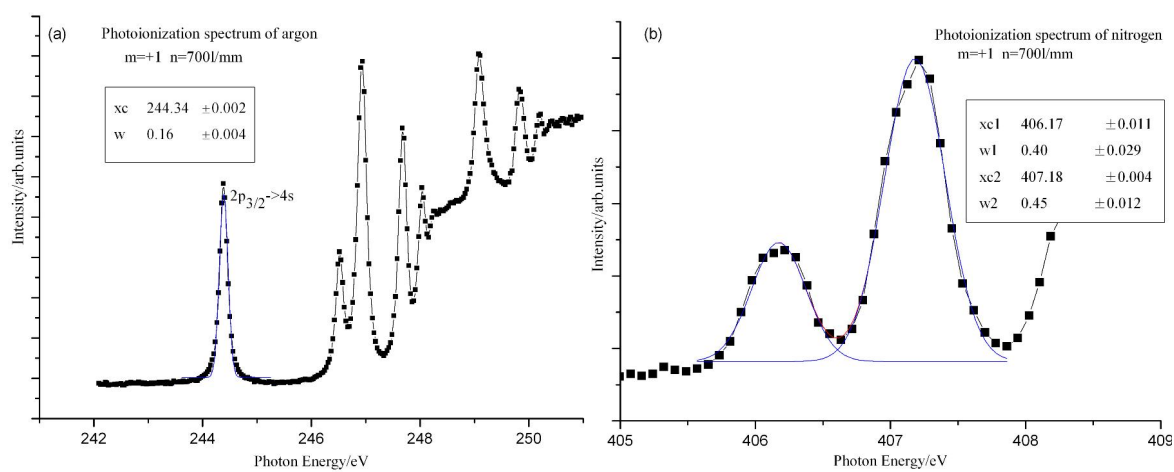

Fig. 3 Photoionization spectrum of argon (a) and nitrogen (b) with the 700 l/mm grating

Fig.4 (a) shows the photoionization spectrum of argon with the 200 lines/mm grating. The diffraction order is +2. The gas pressure in the portable ionization chamber is set to be 4.1 Pa. FWHM at 122 eV is about 0.17 eV. The resolving power for this spectral line is about 1000. Fig.4 (b) shows the photoionization spectrum of krypton with the 200 lines/mm grating. The gas pressure is 5 Pa. FWHM at 91 eV is 0.115 eV. Assuming a natural linewidth of $\Gamma=0.084$ eV, the resolving power at 91.2 eV is about 1200.

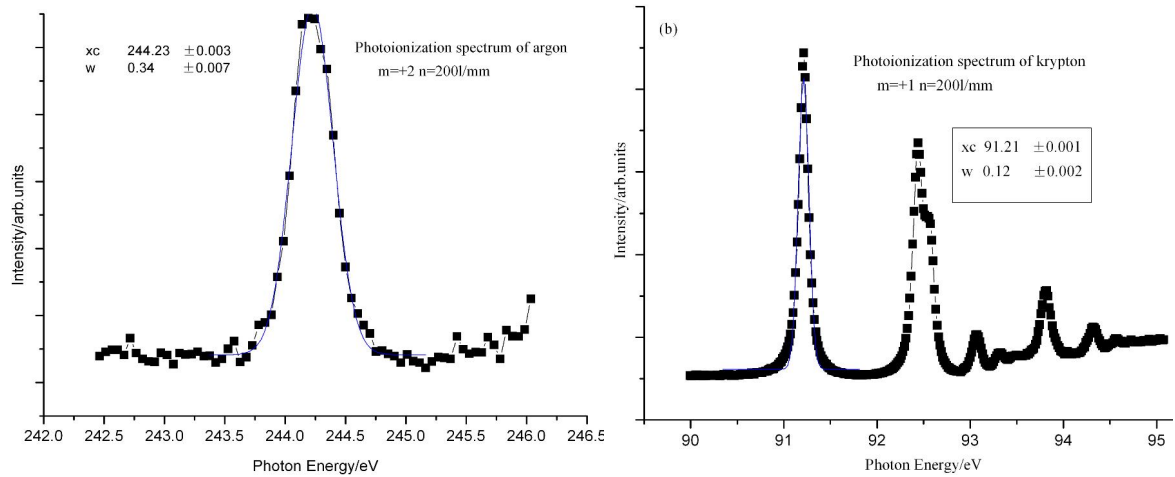

Fig. 4 Photoionization spectrum of argon (a) and krypton (b) with the 200 l/mm grating

### 3.2 Photon Flux

After the measurement of the resolution, the ionization chamber is removed permanently and the sample position is connected with the beamline. Calibrate the photon energies with the photoionization spectrum of gas. Then, the throughput of this beamline is examined with a photodiode. In the testing process, the exit slit is moved along the linear stage to ensuring the focusing of the spectrum. With slits setting of 50 μm and a ring current normalization to 200 mA, the photon flux of the 200 lines/mm grating is found to be better than $10^{10}$ photons/sec/200mA. This meets the original design. However, the photon flux for the 700 lines/mm grating is just $5\times10^{8}$ photons/sec/200mA, and has an obvious reduction around the energy 280eV (the carbon K edge). This phenomenon proves the existence of carbon contamination in this beamline. Fig. 5 shows the results.

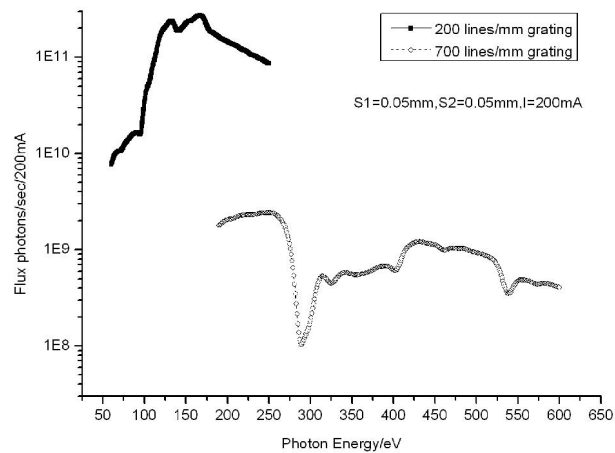

Fig.5 Flux throughput of this beamline

### 3.3 Au 4f photoelectron spectrum

After all the alignment and wavelength calibration, the 4f photoelectron spectrum from atomic Au is tested again to demonstrate the improvement of this beamline. Fig.6 shows the Au 4f, which is measured with 200 eV photons. 700l/mm grating is used, and both entrance and exit slits' widths are set to 100 μm. The total widths of the peaks are 0.57 eV and 0.56 eV for $4f_{5/2}$ and $4f_{7/2}$. The energy difference between these two peaks is 3.68 eV. Assuming a natural linewidth of Γ=0.54 eV, the resolving power E/ΔE is about 1300. This energy resolution is good enough for this beamline. Compared to the spectrum in fig.2, the resolution has a significantly improvement. Table.1 shows the comparison of Au 4f before and after realignment.

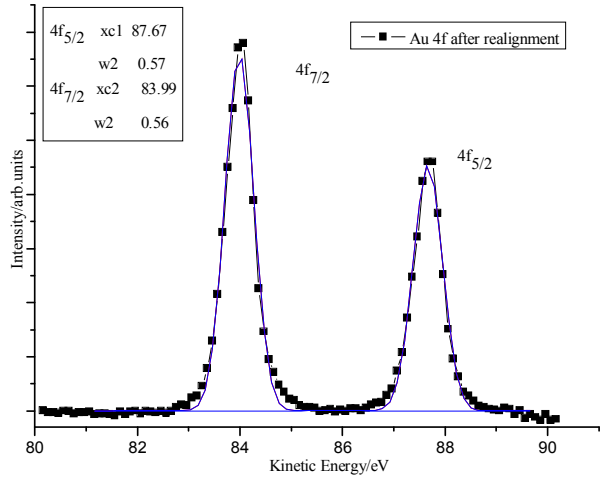

Fig. 6 Au 4f photoelectron spectrum measured with 200 eV photons

Table. 1 Comparison of the FWHM (eV) of Au 4f and energy difference (eV) between these two peaks

|  | Previous | After Realignment |
| --- | --- | --- |
| FWHM (resolving power) of Au $4f_{5/2}$ | 1.21 (410) | 0.57 (1800) |
| FWHM (resolving power) of Au $4f_{7/2}$ | 1.3 (380) | 0.56 (1300) |
| Energy difference between two peaks | 3.62 | 3.68 |

### 4. Conclusion

After realignment, the energy resolving power is up to 1000, which meets the initial design target. The flux of the 200 lines/mm grating is $10^{10}$ photons/s/200mA, which is better than the design value. However, the flux of the 700 lines/mm grating is $5\times10^{8}$ photons/s/200mA, and has an obvious reduction around the carbon K edge, which is caused by carbon contamination. The Au4f photoelectron spectrum after alignment confirms the improvement of the performance.

**Acknowledgements**

The authors acknowledge the help of Xu Xiang-dong in NSRL for his help for the grating test.

# 国家同步辐射实验室光电子能谱光束线调试

李朝阳　潘海斌　魏珅　潘从元　安宁　杜学维　朱俊发　王秋平

中国科学技术大学，国家同步辐射实验室 合肥 230029

**摘要：** 国家同步辐射实验室光电子能谱光束线是使用定包含角单色器的光束线，包含角为 174°。该光束线通过三块线密度分别为 200 线/mm、700 线/mm 以及 1200 线/mm 的球面光栅去覆盖 60eV 到 1000eV 的能量范围。经过多年的使用，光束线的光谱分辨和通量都比较比较低，因此需要对光束线进行重新调试以提高光束线性能。首先对波长扫描机构、光束线中关键部件的姿态以及出缝导轨的姿态进行准直；然后使用原子力显微镜对光栅面型进行检测，最后优化光谱聚焦位置同时使用气体电离室对光束线性能进行测试。光束线的光谱分辨本领达到 1000@244eV。光子通量在使用 200 线/mm 光栅时可达到 $10^{10}$/photons/sec/200mA，在使用 700 线光栅时可达到 $5\times10^{8}$/photons/sec/200mA。由于光学元件表面的碳污染导致 700 线/mm 光栅的通量较低。由于 1200 线/mm 光栅刻槽粗糙度差，光栅衍射效率较低，需要更换一块新的光栅。通过调试，对于 200 线/mm 和 700 线/mm 光栅的光谱分辨率和光子通量都有了很大提高，光束线可以重新投入使用。

**关键字：** 同步辐射；光束线；调试；光通量；光谱分辨率

**PACS:** 07.85.Qe, 41.60.Ap, 42.15.Eq.